\documentclass[11pt,a4paper]{article}
\usepackage{graphicx,color}
\usepackage{amsmath,amssymb,dsfont,fullpage,epsfig,multirow,longtable}
\usepackage{epsfig,epstopdf,hhline}
\usepackage{cases}
\usepackage{wrapfig}
\usepackage{algorithm}
\usepackage{algorithmic}

\usepackage[numbers]{natbib}
\usepackage[top=1 in, bottom=1 in, left=1 in, right=1 in, letterpaper]{geometry}
\usepackage{tikz}
\usetikzlibrary{arrows,positioning,decorations.pathreplacing,shapes}

\usepackage{thmtools}
\usepackage{thm-restate}
\usepackage{enumerate}

\usepackage{enumerate}
\usepackage{mdframed}
\usepackage{setspace}
\usepackage{authblk}

\newenvironment{proof}{\noindent\emph{Proof\ }}{\hspace*{\fill}$\Box$\medskip}

\newtheorem{theorem}{Theorem}

\newtheorem{lemma}{Lemma}
\newtheorem{claim}{Claim}

\newtheorem{corollary}{Corollary}
\usepackage{amsmath}
\usepackage{paralist}
\usepackage{framed}

\newcommand\restr[2]{{
  \left.\kern-\nulldelimiterspace 
  #1 
  \vphantom{\big|} 
  \right|_{#2} 
  }}

\newcommand{\vect}[1]{\ensuremath{\mathbf{#1}}}

\usepackage[utf8]{inputenc} 
\usepackage[T5]{fontenc} 

\usepackage{color, colortbl}
\usepackage{hyperref}
\hypersetup{colorlinks,
            linkcolor=blue,
            citecolor=blue,
            urlcolor=magenta,
            linktocpage,
            plainpages=false}
\usepackage[capitalise,noabbrev]{cleveref}

\begin{document}

\title{Online Covering with Multiple Experts}

\author{Kevi Enik\H{o}}
\author{Nguyễn Kim Thắng}
\affil{LIG, University Grenoble-Alpes, France }

\maketitle

\begin{abstract}
  Designing online algorithms with machine learning predictions is a recent technique beyond the worst-case paradigm for various practically relevant online problems (scheduling, caching, clustering, ski rental, etc.). While most previous learning-augmented algorithm approaches focus on integrating the predictions of a single oracle,
  we study the design of online algorithms with \emph{multiple} experts. To go beyond the popular benchmark of a static best expert in hindsight, we propose a new \emph{dynamic} benchmark (linear combinations of predictions that change over time).
  We present a competitive algorithm in the new dynamic benchmark with a performance guarantee of $O(\log K)$, where $K$ is the number of experts,
  for $0-1$ online optimization problems. Furthermore, our multiple-expert approach provides a new perspective on how to combine in an online manner several online algorithms - a long-standing central subject in the online algorithm research community.
\end{abstract}

\section{Introduction}


The domain of algorithms with predictions \cite{MitzenmacherVassilvitskii20:Beyond-the-Worst-Case} - or learning augmented algorithms - emerged recently and grown immensely at the intersection of (discrete) algorithm design and machine learning (ML).
Combining ML techniques with traditional algorithm design methods enables online algorithms to benefit from predictions that can infer future information from patterns in past data. Online algorithms with predictions can obtain performance guarantees beyond the worst-case analysis and provide fine-tuned solutions to various problems. In the literature, many significant problems have new learning-augmented results, for example, scheduling \cite{LattanziLavastida20:Online-scheduling,Mitzenmacher20:Scheduling-with}, caching (paging) \cite{LykourisVassilvtiskii18:Competitive-caching,Rohatgi20:Near-optimal-bounds,AntoniadisCoester20:Online-metric}, ski rental \cite{GollapudiPanigrahi19:Online-algorithms,KumarPurohit18:Improving-online,AngelopoulosDurr20:Online-Computation}, counting sketches \cite{HsuIndyk19:Learning-Based-Frequency}, bloom filters \cite{KraskaBeutel18:The-case-for-learned,Mitzenmacher18:A-model-for-learned}, and metric task systems \cite{AntoniosEtAll23:mixing-predictions-metric-algorithms}.

Even though predictions provide a glimpse of the future, there is no mathematical guarantee of their accuracy. Adjusting the algorithm's trust in the predictions is a significant challenge since online algorithms must make irrevocable decisions at each time step. Ideally, if the predictions are accurate, the algorithm should perform well compared to the offline setting. In contrast, if the predictions are misleading, the algorithm should maintain a competitive solution, similar to the online setting where no predictive information is available. In other words, online algorithms with predictions are expected to bring the best of both worlds: mathematical performance guarantees of classical algorithms and good future prediction capabilities of machine learning methods.

Predictions can come from multiple sources (heuristics, oracles, randomized methods, etc.), but we ignore their nature and call all of them \emph{experts}.  An algorithm's consistency with the experts' suggestions is typically measured by comparing the algorithm's result with the solution of the \emph{best} expert. A representative example is the popular notion of regret in online learning, which fueled the development of many powerful algorithms and techniques.

A natural research question is whether it is possible to design competitive algorithms with mathematical performance guarantees with a stronger benchmark than the best expert. Comparing an algorithm with a stronger benchmark could provide deeper insights into the learning process and give better ways of exploiting the experts' predictions.

Taking a broader view, we can study whether combining predictions of several experts is similar to combining multiple online algorithms and whether we can expect to achieve better solutions with the combination. Assuming that we do not know in advance which of the given algorithms would perform best on the upcoming requests, can we combine the algorithms in some generic way to obtain a competitive online strategy? This has been a long-standing question in the community of online algorithms \cite{AzarBroder93:On-line-Choice,BlumBurch00:On-line-Learning}. To find an answer, it is a crucial to understand to what extent such an online strategy can benefit from the input of multiple algorithms and what is a suitable benchmark to evaluate its performance.

While in a completely general setting such an online strategy and a corresponding benchmark may not exist, in our paper we propose
an algorithm for online linear problems with covering constraints that is competitive with a new benchmark (informally the \emph{best linear combination} of the experts). Therefore, our paper partially addresses the question we raised in the previous paragraph.

\subsection{Model and Problem}

\paragraph{Covering problem with experts.}
We have $n$ resources and each resource $i$ has a cost per unit $c_{i}$ that we know in advance ($1 \leq i \leq n$).
Let $x_{i}$ be a non-negative variable representing the amount chosen from resource $i$.
The total cost of a solution $(x_{i})_{i=1}^{n}$ is $\sum_{i=1}^{n} c_{i} x_{i}$.
The problem includes $K$ experts and the problem's (covering type) constraints are revealed online (one by one).
At each time $t \geq 1$, we receive a covering constraint $\sum_{i=1}^{n} a_{i}^{t} x_{i} \geq 1$ (where $a_{i}^{t} \geq 0$) and each expert $k$ (where $1 \leq k \leq K$) provides
a solution $(s_{i,k}^{t})_{i=1}^{n}$. An algorithm can observe the experts' solutions and afterwards it must update its own solution (denoted as $(x_{i}^{t})_{i=1}^{n}$)
to satisfy the new constraint, while maintaining the satisfaction of the previous ones. This algorithm must update its solution in the sense of online algorithms, so it cannot modify the previously made decisions. Formally, $x_{i}^{t} \geq x_{i}^{t-1} ~\forall\ i, t$.
Our goal is to design such an algorithm and minimize $\sum_{i=1}^{n} c_{i} x_{i}^{T}$ subject to
all online covering constraints $t$, where $1 \leq t \leq T$. The value $T$ is the last time a constraint is released, and it is not known by the algorithm.

\paragraph{Experts.} In our model, the experts' predictions are also online solutions. In other words, the experts' solutions
fulfill the following properties:
\begin{enumerate}
	\item for every expert $k$ and for every time $t$ the solution $(s_{i,k}^{t})_{i=1}^{n}$ is feasible, therefore, every constraint $t'$ where $1 \leq t' \leq t$ is satisfied;
	\item for every expert $k$ and for every time $t$ and for every resource $i$, the previous expert solutions are irrevocable, therefore $s_{i,k}^{t} \geq s_{i,k}^{t'}$ for all $t' \leq t$.
\end{enumerate}
These properties can be verified online. If some experts do not satisfy them, we simply ignore those experts both in the decision-making and in the benchmark.
A crucial remark: we do \emph{not} assume that the experts' solutions must be tight at each constraint $t$, meaning that $\sum_{i=1}^{n} a_{i}^{t} s_{i,k}^{t} = 1 ~ \forall t, k$.
This assumption is unrealistic and cannot be maintained in an online manner (see the discussion in Appendix~\ref{appix-tight-solutions}).
Besides, assuming tight constraint satisfaction would simplify the problem, while intuitively,
the difficulty of designing competitive algorithms comes from the lack of obvious ways to distinguish
good expert solutions from (probably many) non-efficient/misleading ones.

\paragraph{Benchmark.}
We consider a dynamic benchmark that intuitively captures the \emph{best linear combination} of all experts' solutions \emph{over time}.
Informally, at any online time step, the benchmark can take a linear combination of the experts' solutions.
The linear combination can be changed over time, and it can be different from previous combinations.
However, the benchmark's decisions are also online, so it cannot decrease the value of the decision variables ($x_{i}$).
We refer to our benchmark with the name \texttt{LIN-COMB} from now on.

The \texttt{LIN-COMB} benchmark's formal description is a linear program, visible on \cref{fig:benchmark}.
Let $w_{k}^{t} \geq 0$ be the weight assigned by the \texttt{LIN-COMB} benchmark to expert $k$ (where $1 \leq k \leq K$) at time~$t$.
Since we consider a linear combination, the constraint $ \sum_{k=1}^{K} w_{k}^{t} = 1$ must hold.
The solution of \texttt{LIN-COMB} at time $t$ is ideally $x_{i}^{t} = \sum_{k=1}^{K} w_{k}^{t} s_{i,k}^{t}$,
however, $x_{i}^{t}$ must be larger than $x_{i}^{t-1}$.
Therefore, we set $x_{i}^{t} = \max\bigl\{\sum_{k=1}^{K} w_{k}^{t} s_{i,k}^{t},\ x_{i}^{t-1}\bigr\}$.
In other words, given the chosen weights, if  $\sum_{k=1}^{K} w_{k}^{t} s_{i,k}^{t} < x_{i}^{t-1}$ then $x_{i}^{t} \gets x_{i}^{t-1}$,
otherwise $x_{i}^{t} \gets \sum_{k=1}^{K} w_{k}^{t} s_{i,k}^{t}$.

Since every expert's solution is feasible by our assumptions, at each time $t$ and for all resource $i$ (where $1 \leq i \leq n$),
the constructed solution $x_{i}^{t} \geq \sum_{k=1}^{K} w_{k}^{t} s_{i,k}^{t}$ constitutes a feasible solution to the covering constraints of the original covering problem.
Formally, for every constraint $t'$ with $t' \leq t$,
\begin{align*}
\sum_{i=1}^{n} a_{i}^{t'} x_{i}^{t} \geq
\sum_{i=1}^{n} a_{i}^{t'} \biggl( \sum_{k=1}^{K} w_{k}^{t} s_{i,k}^{t} \biggr)
	= \sum_{k=1}^{K} w_{k}^{t}  \biggl( \sum_{i=1}^{n} a_{i}^{t'} s_{i,k}^{t} \biggr)
	\geq \sum_{k=1}^{K} w_{k}^{t} \geq 1
\end{align*}
where the second inequality holds due to the feasibility of the experts' solutions.
%
%
We highlight that the best-expert benchmark is included in \texttt{LIN-COMB}. By setting $w^{t}_{k^{*}} = 1$ for all $t$, where $1 \leq t \leq T$, and $w^{t}_{k} = 0$ for all $k \neq k^{*}$
where $k^{*}$ we get the best expert in hindsight (so $x_{i}^{t} = s_{i,k^{*}}^{t}$ for all $i,t$).

\begin{figure}
\begin{mdframed}
	\begin{align*}
		&& \min \sum_{i=1}^{n} c_{i} x_{i}^{T} &= \min \sum_{i=1}^{n} c_{i} \sum_{t=1}^{T}\bigl( x_{i}^{t} - x_{i}^{t-1}\bigr)\\
	\text{subject to} &&
		\sum_{k=1}^{K} w_{k}^{t} &= 1  && \forall\ t \\
		&& x_{i}^{t} &\geq \sum_{k=1}^{K} w_{k}^{t} s_{i,k}^{t} && \forall\ i, t\\
		&& x_{i}^{t} &\geq x_{i}^{t-1} && \forall\ i, t\\
		&& w_{k}^{t} &\geq 0  && \forall\ t, k
	\end{align*}
	where $1 \leq t \leq T$ and $1 \leq i \leq n$.
	\vspace{5pt}
\end{mdframed}
\caption{Formulation of the \texttt{LIN-COMB} benchmark}
\label{fig:benchmark}
\end{figure}

\subsection{Our approach and contribution}

To design competitive algorithms with the new benchmark, we consider a primal-dual approach. First, we relax the linear program formulation of \texttt{LIN-COMB} (visible on \cref{fig:benchmark}), which serves as a lower bound. Then, we take the dual of the relaxation, which is a lower bound on the relaxation. Therefore, following the chain of lower bounds, the dual problem is a lower bound on the \texttt{LIN-COMB} benchmark.

At every time step during the execution, our algorithm constructs decisions based on the solution of a convex program. Our approach is inspired by the
convex regularization method of \cite{BuchbinderChen14:Competitive-Analysis}. The objective of the convex program is a shifted entropy function.
These functions have been widely used, in particular in the recent breakthrough related to $k$-server \cite{BubeckCohen18:K-server-via-multiscale,BuchbinderGupta19:k-servers-with}
and metrical tasks problems \cite{BubeckCohen21:Metrical-task},
in which the entropy functions are shifted by constant parameters.
A \emph{novel} point in our approach is that the entropy function is shifted by the average of the experts' solutions.
Moreover, regarding the constraints of the convex program, instead of using the experts' solutions directly,
we define auxiliary solutions that guarantee tight constraint satisfaction and use them in the constraints.
Intuitively, this eliminates/modifies the malicious experts' solutions.

Let $\rho$ be the maximum ratio between the experts' solutions on the resources. Formally,
\[
	\rho := \max_{i} \max_{t',t''} \biggl\{\frac{\sum_{k=1}^{K} s_{i,k}^{t'}}{\sum_{k=1}^{K} s_{i,k}^{t''}} \biggr\}  \textnormal{ s.t. } \sum_{k=1}^{K} s_{i,k}^{t''} > 0.
\]
Informally, $\rho$ represents the discrepancy across the experts' predictions.
Our main result is an algorithm that has an objective cost at most $O(\ln(K\rho))$ times the cost of the \texttt{LIN-COMB} benchmark.
In particular, for $0$-$1$ optimization problems, where the experts provide integer (deterministic or randomized) solutions, our algorithm is $O(\ln K)$-competitive with \texttt{LIN-COMB}.
An interesting feature of our algorithm is its resilience to the fluctuation of the quality of predictions (as discussed in the section below and illustrated in the experiments).

\subsection{Related work and discussions}

\begin{wrapfigure}{r}{0.4\textwidth}
 \vspace{-0.8cm}
  \begin{center}
    \includegraphics[width=0.4\textwidth]{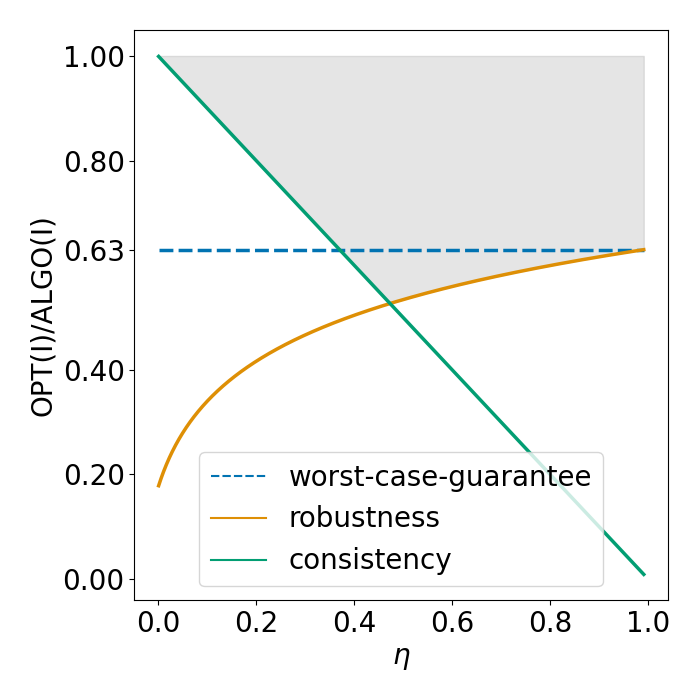}
  \end{center}
  \vspace{-1cm}
  \caption{Robustness-Consistency}
  \label{fig:robustness-consistency}
   \vspace{-0.5cm}
\end{wrapfigure}

Much of the research focusing on surpassing worst-case performance guarantees is motivated by the spectacular advances of machine learning (ML). Specifically, ML methods can detect patterns among the arriving input requests and provide valuable insights for the online algorithms regarding future requests. \cite{LykourisVassilvtiskii18:Competitive-caching} introduced a general framework to integrate ML predictions into classical algorithm designs to surpass the worst-case performance limit.
As a result, many practically relevant online problems were revisited to enhance existing classical algorithms with ML predictions (see the aforementioned \cite{LattanziLavastida20:Online-scheduling,Mitzenmacher20:Scheduling-with,LykourisVassilvtiskii18:Competitive-caching,Rohatgi20:Near-optimal-bounds,AntoniadisCoester20:Online-metric,GollapudiPanigrahi19:Online-algorithms,KumarPurohit18:Improving-online,AngelopoulosDurr20:Online-Computation,HsuIndyk19:Learning-Based-Frequency,KraskaBeutel18:The-case-for-learned,Mitzenmacher18:A-model-for-learned,AntoniosEtAll23:mixing-predictions-metric-algorithms}).

On a high-level view, we aim to design algorithms that are robust (competitive) to the offline optimal solution and also consistent with the expert's predictions. Ideally, the performance of the designed algorithm should surpass previous bounds whenever the predictions are reliable (low errors).
However, most learning-augmented algorithms suffer when the error rates are neither very low nor very high, resulting in prediction confidence that is neither very low nor very high.
Figure~\ref{fig:robustness-consistency} provides a general picture of the performance of an algorithm with predictions, which is representative for many problems (for example, \cite{BamasMaggoriSvensson20:primal-dual-method,KeviNguyen23:Primal-Dual-Algorithms}). In the figure, $\eta$ indicates the confidence in the predictions (or equivalently the error rate of predictions). The learning-augmented algorithm's performance bound is the maximum value of the green and orange curves (gray shaded area on the figure). We can observe that when $0.4 \leq \eta \leq 0.9$,
the algorithm's performance guarantee is worse than the classical worst-case guarantee (that can be achieved by simply ignoring all predictions).
Intuitively, in the case of neither very low nor very high confidence in the predictions, the algorithm has a hard time deciding if it should follow the predictions or the best-known standard algorithm in the worst-case paradigm.
It naturally raises the question of whether one can surely guarantee to achieve at least a constant factor of the worst-case guarantee (where the constant is as close to 1 as possible), assuring the resilience of the output solutions despite the quality of the predictions.
Our algorithm, together with the new benchmark, provides an answer to this question.

The paper of \cite{AnandGe22:Online-Algorithms} is closely related to ours and studies the design of algorithms with multiple experts.
They consider a \texttt{DYNAMIC} benchmark that is intuitively
the minimum cost solution that is supported by at least one expert solution at each step. Formally:
\[\texttt{DYNAMIC} = \min_{\hat{\textbf{x}} \in \hat{X}} \sum_{i=1}^{n} c_i \hat{x}_i \textnormal{, where}\]
\[\hat{X} = \{\hat{\vect{x}} : \forall\ i \in [n],\ \forall\ t \in [T],\ \exists\ k \in [K]\ \textnormal{ such that } s_{i,k}^{t} \le \hat{x}_i \}\]
Our benchmark, \texttt{LIN-COMB}, is included in \texttt{DYNAMIC}, since every solution $x_{i}^{t}$ in \texttt{LIN-COMB} satisfies:
\[
	x_{i}^{t} \geq \sum_{k} s_{i,k}^{t}w_{k}^{t} \geq \min_{k} \{s_{i,k}^{t}\}
\]
therefore, for any $i$ and $t$, there exists $k$ such that $x_{i}^{t} \geq s_{i,k}^{t}$.
However, the inverse is not true: a solution $\hat{\vect{x}}^{t} \in \hat{X}$ in \texttt{DYNAMIC} is not necessarily
a linear combination of the experts' solutions.
The \texttt{DYNAMIC} benchmark in \cite{AnandGe22:Online-Algorithms} relied on the assumption that at every time step
the experts' solutions are tight. This assumption is unrealistic and impossible to maintain in online solutions (see Appendix~\ref{appix-tight-solutions}).
\cite{AnandGe22:Online-Algorithms} claimed an $O(\log K)$-competitive algorithm in the \texttt{DYNAMIC} benchmark.
Unfortunately, this is \emph{incorrect}; we show an example in Appendix~\ref{sec:counter-example}
in which their algorithm's performance guarantee is unbounded in the \texttt{DYNAMIC}
benchmark.

Integrating multiple predictions into the online algorithm design was a topic of other papers as well.
As an example, \cite{GollapudiPanigrahi19:skirental-multiple-predictions} studied the ski rental problem with multiple predictions.
The authors defined a consistency metric, which compares the performance of their algorithm to the optimal solution, given that at least one prediction (among the $k$ predictions) is optimal.
\cite{AlmanzaChierichetti21:Online-Facility} also considered multiple predictions in the online facility location problem.
They compared the performance of their algorithm to the best possible solution obtained on the union of the suggestions. Recently, \cite{DinitzIm:Algorithms-with} studied the use of multiple predictors for several problems such as matching, load balancing, and non-clairvoyant scheduling. They provided algorithms competitive to the best predictor for such problems.
An important remark: all the above benchmarks are captured within \texttt{LIN-COMB}.

Furthermore, \cite{AntoniosEtAll23:mixing-predictions-metric-algorithms} proposed an algorithm with multiple experts for the metrical task system problem. Their benchmark allows switching from one expert to another at each time step, but it does not allow combinations of experts or any solution not suggested by one of the experts. In our \texttt{LIN-COMB} benchmark, the linear combinations that evolve over time could result in a solution that is not suggested by one of the experts and potentially they can be much more efficient. In \cite{AntoniosEtAll23:mixing-predictions-metric-algorithms} there is a cost for state transitions, which is appropriate for their problems, but in many other problems, the smooth transition with additional costs from previous decisions to new ones is not allowed (past decisions are immutable). Therefore, the results of \cite{AntoniosEtAll23:mixing-predictions-metric-algorithms} are not applicable to our setting.

Combining online algorithms into a new algorithm to achieve better results than the individual input algorithms has been a long-standing online algorithm design question \cite{AzarBroder93:On-line-Choice,BlumBurch00:On-line-Learning}.
Its intrinsic difficulty is similar to the issue we mentioned earlier: when the performance of the given input algorithms (or heuristics) is unclear (especially in the online setting), it is challenging to create a combination that can surpass the performance of the included algorithms.
Following the current development of online algorithm design techniques with multiple predictions, this subject has been renewed with different machine learning approaches. Our paper contributes to this line of research.

\section{Online covering with multiple experts}	\label{sec:covering}

Our proposed algorithm solves online covering problems by creating linear combinations of the solutions proposed by $K$ experts in an online manner.
Recall that we evaluate the performance of our algorithm with the \texttt{LIN-COMB} benchmark (formalized on \cref{fig:benchmark}), which consists of the best linear combination of the experts' solution at each step.

Since our \texttt{LIN-COMB} benchmark is a linear combination of the experts' solutions, the equality $ \sum_{k=1}^{K} w_{k}^{t} = 1$ must hold, where $w_{k}^{t} \geq 0$ is the weight assigned to expert $k$ (where $1 \leq k \leq K$) at time $t$. In the following, we formulate a relaxed version of the \texttt{LIN-COMB} formulation, where
$\sum_{k=1}^{K} w_{k}^{t} \geq 1$. Additionally, the relaxed formulation enables us to avoid the (online) hard constraint requiring $w_{k}^{t} s_{i,k}^{t} \geq w_{k}^{t-1} s_{i,k}^{t-1}$ to hold, and instead, we introduce a new variable, $y_{i}^{t}$, to represent the increase of $x_{i}^{t}$ compared to $x_{i}^{t-1}$. When $w_{k}^{t} s_{i,k}^{t} < w_{k}^{t-1} s_{i,k}^{t-1}$ during the execution, we set the contribution of $i$ at time $t$ to be 0, and therefore, $y_{i}^{t} = 0$.
The relaxed formulation is visible on Figure~\ref{fig:relaxation}.

Due to the relaxed constraint, the optimal solution of the relaxed linear program is a lower bound of our \texttt{LIN-COMB} benchmark. The dual of the relaxation is displayed on Figure~\ref{fig:dual}.

\begin{figure}[ht]
	\begin{mdframed}
		\begin{align*}
			&& \min \sum_{t = 1}^{T} \sum_{i=1}^{n} & c_i y_i^t \\
			(\alpha^{t}) \qquad && \sum_{k=1}^{K} w_{k}^{t} & \geq 1  & \forall\ t \\
			(\beta_{i}^{t}) \qquad && \sum_{k=1}^{K} \left(w_{k}^{t} s_{i,k}^{t} - w_{k}^{t-1} s_{i,k}^{t-1} \right) &\leq y_i^t  &\forall\ i,t\\
			&& w_{k}^{t},\ y_{i}^{t} & \ge 0 & \forall\ i,t,k
		\end{align*}
	\end{mdframed}
	\caption{Formulation of the relaxation of the \texttt{LIN-COMB} benchmark}
	\label{fig:relaxation}
\end{figure}

\begin{figure}[ht]
	\begin{mdframed}
		\begin{align*}
			&& \max \sum_{t=1}^{T} & \alpha^{t} \\
			(x_{k}^{t}) \qquad && \alpha^{t} + \sum_{i=1}^{n} s_{i,k}^{t} ( \beta_{i}^{t+1} - \beta_{i}^{t})   &\leq 0  &\forall\ k,t\\
			(y_{i}^{t}) \qquad && \beta_{i}^{t}   &\leq c_{i}  &\forall\ i,t \\
			&& \alpha_{i}^{t},\ \beta_{i}^{t} & \ge 0 & \forall\ i,t
		\end{align*}
	\end{mdframed}
	\caption{Dual formulation of the relaxation of the \texttt{LIN-COMB} benchmark}
	\label{fig:dual}
\end{figure}

According to the theorem of weak duality, any feasible solution of the dual program lower bounds any feasible solution of the primal program, and therefore, any feasible dual solution also lower bounds our \texttt{LIN-COMB} benchmark. Following the chain of lower bounds, our approach to design a competitive algorithm is as follows. At every time step $t$, we build solutions for all $x_{i}^{t}$ together with the solutions for the dual problem $(\alpha^{t}, \beta_{i}^{t})$. Then, we bound the cost of the algorithm to that of the dual. It is important to emphasize that the designed solution for every $x_{i}^{t}$ must be feasible to the covering constraints, but it may \emph{not necessarily} be a linear combination of the experts' solutions.

\subsection{Competitive Algorithm} \label{sec:algo}

\paragraph{Preprocessing.}
Recall that by our assumptions, the experts' solutions are always feasible and non-decreasing. At the arrival of the $t^{\text{th}}$ constraint, expert $k$ (where $1 \leq k \leq K$) provides a feasible solution $s_{k}^{t} = (s_{i,k}^{t})_{i=1}^{n}$, such that $s_{i,k}^{t} \ge s_{i,k}^{t'}$ for all $t' \le t$ and all $i$ where $1 \le i \le n$. These assumptions do not exclude the possibility for the experts to provide malicious solutions that instruct the algorithm to use an unnecessarily large amount of resources.
Note that contrary to the assumption in \cite{AnandGe22:Online-Algorithms}, we can \emph{not} expect the experts' solutions to be always tight.
(In Appendix~\ref{appix-tight-solutions}
we show an example that tight solutions cannot be maintained in an online manner.)

To circumvent this issue, we preprocess the experts' solutions at each iteration. During the preprocessing, every solution $s_k^t$ is scaled down to make it as tight as possible on the $t^{\text{th}}$ constraint, while always maintaining $s_{i,k}^{t} \geq s_{i,k}^{t-1}$ for all $i$. Additionally, after the down-scaling, we create an auxiliary solution $\hat{s}_k^t$ that is tight for the $t^{\text{th}}$ constraint. This solution is useful for our algorithm, and we create it with the following procedure.

After the down-scaling, do the following for each expert $k$.
\begin{compactenum}
	\item If $(s_{i,k}^{t})_{i=1}^{n}$ is tight on the $t^{\text{th}}$ constraint, then set $\hat{s}_{i,k}^{t} \gets s_{i,k}^{t}$  for every $i$.
	\item Let $\hat{s}_{i,k}^{t-1}$ be the auxiliary solution of expert $k$ at time $t-1$, meaning that, $\sum_{i=1}^{n} a_{i}^{t-1} \hat{s}_{i,k}^{t-1} = 1$. Given $I := \{i: s_{i,k}^{t} > \hat{s}_{i,k}^{t-1} \cdot \frac{a_{i}^{t-1}}{a_{i}^{t}} \}$, we set $\hat{s}_{i,k}^{t} \gets s_{i,k}^{t}$ if $i \notin I$
	and set $\hat{s}_{i,k}^{t}$ to be some value in $[\hat{s}_{i,k}^{t-1} \cdot \frac{a_{i}^{t-1}}{a_{i}^{t}}, s_{i,k}^{t}]$ if $i \in I$, s.t. the solution $\hat{s}_{i,k}^{t}$
	becomes tight on the $t^{\text{th}}$ constraint.
\end{compactenum}
\begin{lemma}
Following the preprocessing procedure, we can always obtain the solutions $\hat{s}_{i,k}^{t}$ such that
$\hat{s}_{i,k}^{t} \leq s_{i,k}^{t}$ and $\sum_{i=1}^{n} a_{i}^{t} \hat{s}_{i,k}^{t} = 1$.
\end{lemma}
\begin{proof}
Let us fix an expert $k$. We prove the lemma by induction. At time step $t=1$, one can always scale down the solution $s_{i,k}^{1} \geq 0$ such that the first constraint becomes tight.
Assume that the lemma holds until $t-1$, $\sum_{i=1}^{n} a_{i}^{t-1} \hat{s}_{i,k}^{t-1} = 1$ and $\hat{s}_{i,k}^{t-1} \leq s_{i,k}^{t-1}$.
Consider time $t$. If after scaling down (at the first step in the procedure) the $t^{\text{th}}$ constraint becomes tight, then we are done. Otherwise, we have
	\begin{align*}
		1 &< \sum_{i=1}^{n} a_{i}^{t} s_{i,k}^{t} = \sum_{i \in I} a_{i}^{t} s_{i,k}^{t} + \sum_{i \notin I} a_{i}^{t} s_{i,k}^{t}, \\
		1 &= \sum_{i=1}^{n} a_{i}^{t-1} \hat{s}_{i,k}^{t-1} =  \sum_{i = 1}^{n} a_{i}^{t} \biggl( \hat{s}_{i,k}^{t-1} \cdot \frac{a_{i}^{t-1}}{a_{i}^{t}} \biggr) \\
		&\geq  \sum_{i \in I} a_{i}^{t} \biggl( \hat{s}_{i,k}^{t-1} \cdot \frac{a_{i}^{t-1}}{a_{i}^{t}} \biggr)
		+ \sum_{i \notin I} a_{i}^{t} s_{i,k}^{t}
	\end{align*}
	Hence, there exists $\hat{s}_{i,k}^{t} \in \bigl[ \hat{s}_{i,k}^{t-1} \cdot \frac{a_{i}^{t-1}}{a_{i}^{t}}, s_{i,k}^{t} \bigr]$ for every $i$, where $1 \leq i \leq n$, such that $\sum_{i=1}^{n} a_{i}^{t} \hat{s}_{i,k}^{t} = 1$.
\end{proof}

\clearpage

\paragraph{Algorithm.}

At the arrival of the $t^{\text{th}}$ constraint,
\begin{compactenum}
	\item solve the following convex program and set $w^t$ to be the obtained optimal solution
\begin{align*}
&& \min_{w} \biggl\{\sum_{i=1}^{n} c_{i}  \biggl[  \biggl(\sum_{k=1}^{K} s_{i,k}^{t} w_{i,k}  + \delta_{i}^{t} \biggr) &
					 \ln \left( \frac{\sum_{k=1}^{K} s_{i,k}^{t} w_{i,k}  + \delta_{i}^{t}}{ \sum_{k=1}^{K}  s_{ik}^{t-1}w_{i,k}^{t-1}  + \delta_{i}^{t-1}}  \right)
					 		- \sum_{k=1}^{K}  s_{i,k}^{t} w_{i,k} \biggr] \biggr\} \\
    (\gamma^{t})  && \sum_{i=1}^{n} a_{i}^{t} \biggl( \sum_{k=1}^{K}  \hat{s}_{i,k}^{t} w_{i,k} \biggr) &\geq 1 \qquad \forall\ t\\
    (\lambda_{i}^{t}) && \sum_{k=1}^{K}  w_{i,k} &\geq 1 \qquad \forall\ i\\
    (\mu_{i}^{t}) && \sum_{k=1}^{K} s_{i,k}^{t} w_{i,k} &\geq 0 \qquad \forall\ i,t
\end{align*}
where $\delta_{i}^{t} = \frac{1}{K} \sum_{k} s_{i,k}^{t}$.
Note that in this program, we use the auxiliary solution $\hat{s}_{i,k}^{t}$ in the first constraint. For every $i$ where $s_{i,k}^{t} = 0$ for all $k$, the term related to $i$ is not included in the objective function of the convex program.
(We can set $w_{i,k} = 0$ for all $k$ beforehand.)
	\item For all $i$ if $\sum_{k=1}^{K} w_{i,k}^{t} s_{i,k}^{t} > x_{i}^{t-1}$ then set $x_{i}^{t} \gets \sum_{k=1}^{K} w_{i,k}^{t} s_{i,k}^{t}$;
otherwise set $x_{i}^{t} \gets x_{i}^{t-1}$.
\end{compactenum}

\medskip

\noindent \textit{Note:} To avoid the possible division by 0 in the denominator of the objective function's logarithm, we can use a dummy expert, who sets each variable to some small value and then follows the greedy heuristic to solve the problem at each arriving constraint. The presence of this expert only changes the competitive ratio to $O(log K + 1)$. Additionally, upon the arrival of the first constraint, we treat the denominator as $1$.

\subsection{Analysis}
As $w^{t}$ is the optimal solution of the convex program and ($\gamma^t,\ \lambda_{i}^{t},\ \mu_{i}^{t}$) is the optimal solution of its dual, the following Karush-Kuhn-Tucker (KKT) and complementary slackness conditions hold.

\begin{align*}
   \biggl[ \sum_{i=1}^{n} a_{i}^{t} \biggl( \sum_{k}  \hat{s}_{i,k}^{t} w_{i,k}^{t} \biggr) - 1 \biggr] \gamma^{t} &= 0 \qquad \forall t \\
   \biggl[ \sum_{k=1}^{K}  w_{i,k}^{t}  - 1 \biggr] \lambda_{i}^{t} &= 0 \qquad \forall i, t \\
   \biggl[ \sum_{k=1}^{K}  s_{i,k}^{t} w_{i,k}^{t} \biggr] \mu_{i}^{t} &= 0 \qquad \forall i, t \\
 c_{i} s_{ik}^{t} \ln \left( \frac{\sum_{k=1}^{K} s_{i,k}^{t} w_{i,k}^{t} + \delta_{i}^{t}}{\sum_{k=1}^{K}  s_{ik}^{t-1}w_{i,k}^{t-1}  + \delta_{i}^{t-1}} \right)
    	- a_{i}^{t} \hat{s}_{i,k}^{t} \gamma^{t} - \lambda_{i}^{t} - s_{i,k}^{t} \mu_{i}^{t} &= 0	\qquad \forall i,k,t \\
	\gamma^{t}, \lambda_{i}^{t}, \mu_{i}^{t} &\geq 0 \qquad \forall i, t
\end{align*}

Moreover, if $\sum_{k=1}^{K} w_{i,k}^{t} s_{i,k}^{t} > 0$, meaning that $\mu_{i}^{t} = 0$, then
\begin{align}	\label{eq:KKT}
   c_{i} s_{ik}^{t} \ln \left( \frac{\sum_{k=1}^{K} s_{i,k}^{t} w_{i,k}^{t}  + \delta_{i}^{t}}{\sum_{k=1}^{K}  s_{ik}^{t-1}w_{i,k}^{t-1}  + \delta_{i}^{t-1}} \right)
    	- a_{i}^{t} \hat{s}_{i,k}^{t} \gamma^{t} - \lambda_{i}^{t} = 0
\end{align}

\paragraph{Dual variables and feasibility.} We set the dual variables of the linear program relaxation of our \texttt{LIN-COMB} benchmark based on the dual variables of the convex program used inside the algorithm.
\begin{align*}
    \alpha^{t} &= \frac{1}{\ln(K\rho)}  \biggl( \gamma^{t} + \sum_{i} \lambda_{i}^{t} \biggr), \\
    \beta_{i}^{t} &= \frac{1}{\ln(K\rho)} c_i \ln \left(\frac{ (1 + 1/K) \cdot \max_{t'} \sum_{k=1}^{K} s_{i,k}^{t'}}{\sum_{k=1}^{K}  s_{i,k}^{t-1} w_{i,k}^{t-1} + \delta_{i}^{t-1}}\right)
\end{align*}
where recall that $\rho = \max_{i, t',t''} \left\{\frac{\sum_{k=1}^{K} s_{i,k}^{t'}}{\sum_{k=1}^{K} s_{i,k}^{t''}} : \sum_{k=1}^{K} s_{i,k}^{t''} > 0 \right\}$.

\begin{lemma} \label{lem:covering-feasibility}
The $x_{i}^{t}$ solutions set by the algorithm for the original covering problem and the dual variables $(\alpha^{t}, \beta_{i}^{t})$ of the \texttt{LIN-COMB} benchmark's linear program relaxation are feasible.
\end{lemma}
\begin{proof}
We first prove that the $x_{i}^{t}$ variables satisfy the covering constraints by induction. At time 0, no constraint has been released yet, and every variable is set to 0. This all-zero solution is feasible. Let us assume that the algorithm provides feasible solutions up to time $t-1$. At time $t$, the algorithm maintains the inequality $x_{i}^{t} \geq x_{i}^{t-1}$, so all constraints $t'$ where $t' < t$ are satisfied. Besides, $x_{i}^{t}$ is always at least
$\sum_{k} w_{i,k}^{t} s_{i,k}^{t}$, which is larger than $\sum_{k} w_{i,k}^{t} \hat{s}_{i,k}^{t}$ since $s_{i,k}^{t} \geq \hat{s}_{i,k}^{t}$
for all $i,k$ by the preprocessing step. Hence, the constraint $t$ is also satisfied, formally,
$$
\sum_{i=1}^{n} a_{i}^{t} x_{i}^{t}  \geq \sum_{i=1}^{n} a_{i}^{t} \biggl( \sum_{k} \hat{s}_{i,k}^{t} w_{i,k}^{t} \biggr) \geq 1.
$$

In the remaining part of the proof, we show the feasibility of $\alpha^{t}$ and every $\beta_{i}^{t}$.
Since $ \gamma^{t} \geq 0$ and $\lambda_{i}^{t} \geq 0$ for all $i$ and $t$, we get that $\alpha^{t} \geq 0$.
In the definition of  $\beta_{i}^{t}$, the nominator of the logarithm term is always larger than the denominator, and it is smaller than $K\rho$ times the denominator. Consequently, $0 \leq \beta_{i}^{t} \leq c_{i}$. Furthermore,
\begin{align*}
    \beta_{i}^{t+1} - \beta_{i}^{t}
    	&= - \frac{1}{\ln(K\rho)} c_i \ln \left( \frac{\sum_{k=1}^{K}  s_{i,k}^{t} w_{i,k}^{t} + \delta_{i}^{t}}{\sum_{k=1}^{K}  s_{i,k}^{t-1}w_{i,k}^{t-1} + \delta_{i}^{t-1}} \right).
\end{align*}
Since $\sum_{i} a_{i}^{t} \hat{s}_{i,k}^{t} = 1$, using the KKT conditions, we get:
\begin{align*}
\alpha^{t} + \sum_{i=1}^{n} s_{ik}^{t} \left(\beta_{i}^{t+1} - \beta_{i}^{t}\right)
&= \frac{1}{\ln(K\rho)} \biggl( \gamma^{t} + \sum_{i} \lambda_{i}^{t} \biggr)
	- \frac{1}{\ln(K\rho)}  \sum_{i=1}^{n} s_{i,k}^{t} c_i \ln \left( \frac{\sum_{k=1}^{K}  s_{ik}^{t} w_{i,k}^{t} + \delta_{i}^{t}}{\sum_{k=1}^{K}  s_{ik}^{t-1}w_{i,k}^{t-1} + \delta_{i}^{t-1}} \right) \\
&= \frac{1}{\ln(K\rho)} \biggl[ \gamma^{t} + \sum_{i=1}^{n} \lambda_{i}^{t} - \sum_{i=1}^{n} \left( a_{i}^{t} \hat{s}_{i,k}^{t} \gamma^{t} + \lambda_{i}^{t} + s_{i,k}^{t} \mu_{i}^{t} \right) \biggr] \\
&\leq 0
\end{align*}
\end{proof}

\clearpage

\begin{theorem} \label{covering-theorem}
The algorithm's cost is at most $O(\ln(K \rho))$-competitive in the \texttt{LIN-COMB} benchmark.
\end{theorem}
\begin{proof} \cref{lem:covering-feasibility} proved that our algorithm creates feasible solutions for the dual problem of the \texttt{LIN-COMB} benchmark relaxation and for the original covering problem. We show that the algorithm's solution increases the primal objective value of the original covering problem by at most $O(\ln(K \rho))$ times the value of the dual solution, which serves as the lower bound on the \texttt{LIN-COMB} benchmark - the best linear combination of the experts' solutions.
\begin{align}
	 \sum_{i=1}^{n} &c_{i} (x_{i}^{t} - x_{i}^{t-1})
		= \sum_{i: x_{i}^{t} > x_{i}^{t-1}} c_{i}(x_{i}^{t} - x_{i}^{t-1}) &&  \notag \\
		&\leq \sum_{i: x_{i}^{t} > x_{i}^{t-1}} c_{i}(x_{i}^{t} + \delta_{i}^{t}) \ln \frac{x_{i}^{t} + \delta_{i}^{t}}{x_{i}^{t-1} + \delta_{i}^{t}} \\
		&\leq \sum_{i: x_{i}^{t} > x_{i}^{t-1}} c_{i} (x_{i}^{t} + \delta_{i}^{t}) \ln \frac{x_{i}^{t} + \delta_{i}^{t}}{x_{i}^{t-1} + \delta_{i}^{t-1}} \\
		&= \sum_{i: x_{i}^{t} > x_{i}^{t-1}} c_{i} \left[ \left(\sum_{k=1}^{K}  s_{i,k}^{t} w_{i,k}^{t} + \frac{1}{K} \sum_{k=1}^{K} s_{i,k}^{t} \right)
			\ln \left(\frac{ \sum_{k=1}^{K}  s_{i,k}^{t} w_{i,k}^{t} + \delta_{i}^{t}}{x_{i}^{t-1} + \delta_{i}^{t-1}}  \right) \right] \\
&\leq \sum_{i: x_{i}^{t} > x_{i}^{t-1}} c_{i} \left[ \left(\sum_{k=1}^{K}  s_{i,k}^{t} w_{i,k}^{t} + \frac{1}{K} \sum_{k=1}^{K} s_{i,k}^{t} \right)
			\ln \left(\frac{ \sum_{k=1}^{K}  s_{i,k}^{t} w_{i,k}^{t} + \delta_{i}^{t}}{\sum_{k=1}^{K}  s_{i,k}^{t-1} w_{i,k}^{t-1} + \delta_{i}^{t-1}}  \right) \right]\\
	&= \sum_{i: x_{i}^{t} > x_{i}^{t-1}} \sum_{k=1}^{K} (w_{i,k}^{t} + 1/K) c_{i} s_{i,k}^{t}
				\ln \left(\frac{ \sum_{k=1}^{K} s_{i,k}^{t} w_{i,k}^{t}  + \delta_{i}^{t}}{\sum_{k=1}^{K}  s_{i,k}^{t-1} w_{i,k}^{t-1}  + \delta_{i}^{t-1}}  \right) \notag \\
%
%
&=  \sum_{i: x_{i}^{t} > x_{i}^{t-1}} \sum_{k=1}^{K} (w_{i,k}^{t} + 1/K) \biggl( a_{i}^{t} \hat{s}_{i,k}^{t} \gamma^t + \lambda_{i}^{t} \biggr) \\
&\leq \sum_{i=1}^{n} \sum_{k=1}^{K} (w_{i,k}^{t} + 1/K) \biggl( a_{i}^{t} \hat{s}_{i,k}^{t} \gamma^t + \lambda_{i}^{t} \biggr) \notag \\
&= \sum_{i=1}^{n} a_{i}^{t} \biggl(\sum_{k=1}^{K} w_{i,k}^{t} \hat{s}_{i,k}^{t} \biggr) \gamma^t + \sum_{i=1}^{n} \bigg( \sum_{k=1}^{K} w_{i,k}^{t} \biggr) \lambda_{i}^{t}
+ \frac{1}{K}  \sum_{k=1}^{K} \biggl( \sum_{i=1}^{n} a_{i}^{t}  \hat{s}_{i,k}^{t}  \biggr) \gamma^t + \frac{1}{K} \sum_{k=1}^{K} \sum_{i=1}^{n} \lambda_{i}^{t} 		\notag \\
&= 2 \gamma^{t} + 2\sum_{i=1}^{n} \lambda_{i}^{t} = \ln(K \rho) \alpha^{t}
\end{align}
The above corresponding transformations hold since:
\begin{compactenum}[(1)]
	\setcounter{enumi}{1}
	\item follows from the inequality $a - b \leq a \ln(a/b)$ for all $0 < b \leq a$;
	\item holds since $\delta_{i}^{t} \geq \delta_{i}^{t-1}$ (because $s_{i,k}^{t} \geq s_{i,k}^{t-1}$ for all $i,k,t$);
	\item is valid because $x_{i}^{t} > x_{i}^{t-1}$, so $x_{i}^{t} = \sum_{k=1}^{K}  s_{i,k}^{t} w_{i,k}^{t}$;
	\item is by the design of the algorithm: $x_{i}^{t-1} \geq \sum_{k=1}^{K}  s_{i,k}^{t-1} w_{i,k}^{t-1}$;
	\setcounter{enumi}{5}
	\item since given that $x_{i}^{t} > x_{i}^{t-1} \geq 0$
	(so $\sum_{k=1}^{K}  s_{i,k}^{t} w_{i,k}^{t} = x_{i}^{t} > 0$), the KKT condition (\ref{eq:KKT}) applies;
	\item is true due to the complementary slackness conditions
		and that $\sum_{i=1}^{n} a_{i}^{t}  \hat{s}_{i,k}^{t} = 1$.
\end{compactenum}
\end{proof}

\begin{corollary} \label{corollary}
	For $0$-$1$ optimization problems in which experts provide integer (deterministic or randomized) solutions,
	the algorithm is $O(\ln K)$-competitive in the \texttt{LIN-COMB} benchmark.
	Subsequently, there exists an algorithm such that its performance is $O(\ln K)$-competitive in the \texttt{LIN-COMB}
	benchmark and is up to a constant factor to the best guarantee in the worst-case benchmark
\end{corollary}
\begin{proof}

\begin{wrapfigure}{r}{0.4\textwidth}
 \vspace{-0.9cm}
  \begin{center}
    \includegraphics[width=0.4\textwidth]{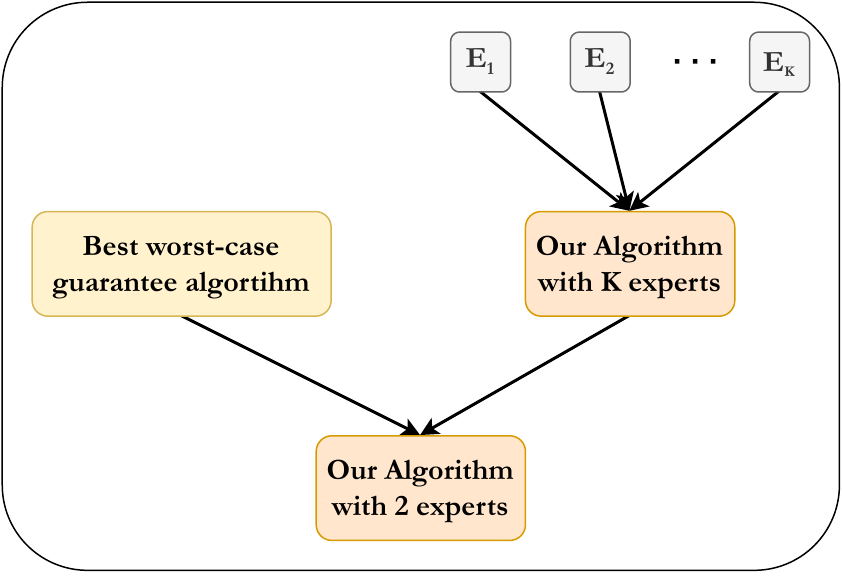}
  \end{center}
  \vspace{-0.5cm}
      \caption{Structural overview of the algorithm's components. $E_1,\ E_2, \dots\ E_K$ correspond to the experts of the online problem.
      On the second layer, we integrate the best standard online algorithm with our algorithm. }
  \label{fig:algo-layers}
   \vspace{-0.5cm}
\end{wrapfigure}

	If the value of $s_{i,k}^{t}$ is in $\{0,1\}$ for every $i,k,t$, then
	\[
	\rho = \max_{i} \max_{t',t''} \left\{\frac{\sum_{k=1}^{K} s_{i,k}^{t'}}{\sum_{k=1}^{K} s_{i,k}^{t''}} : \sum_{k=1}^{K} s_{i,k}^{t''} > 0 \right\}
	\leq \frac{K}{1}
	\]
	Therefore, the competitive ratio of the main algorithm in the \texttt{LIN-COMB} benchmark is $O(\log K \rho) = O(\log K^2)$.

 	To obtain an algorithm that is competitive in both the \texttt{LIN-COMB} and the worst-case benchmarks, we proceed as follows (an illustration in \cref{fig:algo-layers}).
	We first apply the main algorithm on the $K$ experts' predictions to obtain an online algorithm, named $A$.
	Algorithm $A$ is $O(\ln K)$-competitive in the \texttt{LIN-COMB} benchmark. Let $B$ be the algorithm with the best worst-case guarantee.
	One applies the main algorithm one more on two algorithms, $A$ and $B$. The final algorithm is $O(\ln 2)$-competitive to both $A$ and $B$.
	In other words, its performance is $O(\ln K)$-competitive in the \texttt{LIN-COMB} benchmark and is up to a constant factor worse than the best guarantee in the worst-case benchmark.

\end{proof}

By \cref{corollary}, given a $0$-$1$ optimization problem, if there are $K$ deterministic online algorithms, then
we can design an algorithm that has a cost at most $O(\log K)$ times that of the best linear combination of those algorithms at any time.
Similarly, if $K$ given online algorithms are randomized (they output $0$-$1$ solutions with probabilities), then our algorithm
has an expected cost (randomization over the product of the distributions of those solutions) at most $O(\log K)$ times that of
the best linear combination of those algorithms at any time. Many practical problems admit $0$-$1$ solutions, for which our algorithm is of interest.
Consider problems like network design, ski rental, TCP acknowledgement, facility location, etc. Given the fractional solutions constructed by our algorithm,
we can apply existing online rounding schemes to obtain integral solutions for such problems.

\section{Experiments}

\textbf{Implementation.} The first step of our proposed algorithm is to solve a convex program. In the experiments, we approximate the optimal solution of this program using a vanilla Frank-Wolfe implementation. The linear minimization step within Frank-Wolfe is solved with the Gurobi optimizer.

\textbf{Comparison.} The best standard online algorithm for general covering problems without experts is the online multiplicative weight update (MWA) algorithm. In the experiments we compare our algorithm with the MWA algorithm. When a new constraint arrives in the online problem, the MWA algorithm increases each variable $x_i$ in the constraint with a rate of $\frac{a^t_i}{c_i}(x_i + 1/n)$, where $n$ is the total number of variables. We also compare our results with the optimum offline solution (that knows the whole instance in advance) and the average solution of the experts.

\textbf{Input.} First, we evaluated the result of our algorithm on the pathological input of the MWA algorithm. This instance includes $n$ variables and $n$ constraints with uniform costs and coefficients. Each arriving constraint in this pathological example includes one less variable. While the optimal solution is $1$, the worst-case guarantee of MWA is $O(\log n)$. For our algorithm we provided $n$ experts, where $(n-1)$ experts suggest an adversarial trivial solution to set all variables to $1$, while $1$ expert suggests the optimal offline solution. The result of this experiment is visible on \cref{fig:exp-objective}. An important highlight: our algorithm managed to identify the good expert among the majority of adversaries, obtaining a better objective value, than MWA. Second, we experimented with the counter example (in Appendix~\ref{sec:counter-example}) to show that the algorithm proposed by \cite{AnandGe22:Online-Algorithms} has an incorrect performance proof. The result is visible on \cref{fig:exp-objective}. Finally, we generated some instances to observe the performance of our algorithm on non-specific inputs.
The specification for the instance generation includes several parameters, which we detail on \cref{fig:exp-params}.

\textbf{Result.} Multiplicative weight update is a simple and well-performing algorithm in practice. On its pathological worst-case example, our algorithm performs better, however on most instances the expert suggestions were significantly worse than MWA, which impacted the performance of our algorithm. To some extent, our algorithm can detect the good experts and is robust against even many adversaries. We think that given a real-life problem, it is possible to construct well-performing experts (for example using past data) and with a fine-tuned convex program solver our algorithm can be of interest for various use-cases.

\begin{figure}[!ht]
\centering
\begin{tabular}{r|r|r|r|r|r|r}
          & Worst-case & Counter-ex. & & & & \\
Algo name & for MWA  & for \cite{AnandGe22:Online-Algorithms} & Inst. $1$ & Inst. $2$ & Inst. $3$ & Inst. $4$\\
\hline
OPT Offline            & 1.0 & 1.0 & 1.3 & 1.5 & 10.6 & 31.3 \\
MWA Online             & 2.9 & 2.3 & 2.0 & 1.7 & 28.1 & 63.7 \\
\hline
Our Algo           & {\bf 2.2} & {\bf 4.4} & {\bf 2.0} & {\bf 1.7} & {\bf 26.7}  & {\bf 61.7} \\
\hline
Avg of experts      & 9.1 & 3.5 & 16.4 & 23.3 & 1897.47 & 96.9 \\
\end{tabular}
\caption{Objective value of the experiment instances}
\label{fig:exp-objective}
\end{figure}

\begin{figure}[!ht]
\centering
\begin{tabular}{l|r|r|r|r}
Input Generation Parameters & Instance $1$ & Instance $2$ & Instance $3$ & Instance $4$ \\
\hline
Number of variables            & 10 & 10 & 44  & 30 \\
Number of constraints          & 10 & 25 & 2   & 15 \\
Min objective coefficient      & 1  & 10 & 1   & 1  \\
Max objective coefficient      & 10 & 25 & 100 & 100\\
Min constraint coefficient     & 1  & 10 & 1   & 1 \\
Max constraint coefficient     & 10 & 25 & 1   & 1 \\
Min number of zero coefficient & 0  &  1 & 11  & 5 \\
Max number of zero coefficient & 5  &  5 & 22  & 20 \\
Number of perfect experts      & 1  &  0 & 0   & 2 \\
Number of online experts       & 2  &  1 & 1   & 2 \\
Number of random experts       & 1  &  1 & 11  & 0 \\
Number of adversaries          & 1  &  1 & 0   & 0 \\
\end{tabular}
\caption{Parameters of the generated experiment instances}
\label{fig:exp-params}
\end{figure}

\section{Conclusion}

We introduce a dynamic \texttt{LIN-COMB} benchmark in the setting of multiple expert predictions beyond the traditional static benchmark of the best expert in hindsight
and give a competitive algorithm for the online covering problem in this benchmark.
Our approach can provide valuable insights into the learning processes related to predictions,
in particular, in aggregating information from predictions to improve the performance of existing algorithms,
and how to combine online algorithms, an important subject in the online algorithm design community \cite{AzarBroder93:On-line-Choice,BlumBurch00:On-line-Learning}.
The experiments support the fact that our algorithm can differentiate between good and adversarial experts to some extent.

An interesting open question is to design competitive algorithms in the \texttt{LIN-COMB} benchmark for different classes of problems,
such as packing problems and problems with non-linear objectives.

\clearpage

\bibliographystyle{plainurl}
\bibliography{references}

\clearpage

\appendix

\section*{Appendix}
\section{Counter example for tight online expert solutions} \label{appix-tight-solutions}

The following example shows that we cannot expect online expert solutions (in the sense of online algorithms)
to be tight on the arriving constraints. In the example below, we display the experts' solutions after each constraint.

\begin{align*}
    \min\ \ \ \ x_{1} + \ \ \ x_{2} & \\
    x_{1} + \frac{1}{2}\ x_{2} & \ge 1 \\
    \textnormal{Expert}_{1}: \ \ \ \ \ \ \ \ \ \ \ 1 \ \ \ \ \ \ \ \ 0 & \\
    \textnormal{Expert}_{2}: \ \ \ \ \ \ \ \ \ \ \ 0 \ \ \ \ \ \ \ \ 2 & \\
    x_{2} & \ge 1 \\
    \textnormal{Expert}_{1}: \ \ \ \ \ \ \ \ \ \ \ 1 \ \ \ \ \ \ \ \ 1 & \\
    {\color{red}\textnormal{Expert}_{2}:} \ \ \ \ \ \ \ \ \ \ \ \ {\color{red} 0} \ \ \ \ \ \ \ \ {\color{red} 2} & \\
\end{align*}
To have tight a suggestion from Expert$_2$ on the second constraint, Expert$_2$ not only has to decrease its value of $x_{2}$ (which is not allowed), but even increase the value of $x_{1}$ for the first constraint. In other words, Expert$_2$ has to completely modify its past decisions.

\section{Counter example for the performance of the algorithm of \cite{AnandGe22:Online-Algorithms}}
\label{sec:counter-example}

Anand, Ge, Kumar and Panigrahi \cite{AnandGe22:Online-Algorithms} recently proposed online algorithms for online covering problems with multiple expert solutions.
We show here a counter example that contradicts Theorem~$2.1$ presented in Section~$3$ of their paper.
In the proof of Theorem~$2.1$ the authors state that \textit{the total cost of the algorithm is at most $3$ times the potential $\phi$ at the beginning, i.e., at most $O(\log~K)$ times the \texttt{DYNAMIC} benchmark}. However, in our counter example the total cost of their algorithm is $O(L \log(K))$ times the \texttt{DYNAMIC} benchmark, where $L$ is an arbitrary large number.

\subsection{Setting}

Algorithm $1$ (from \cite{AnandGe22:Online-Algorithms}) receives solutions from $K$ experts. The authors denote with $x_i(j,s)$ the solution from expert $s$ for variable $i$ on constraint $j$. They assume that the expert solutions are tight, formally:
\[ \sum_{i=1}^{n} a_{ij}\ x_{i}(j, s) = 1 \ \ \ \ \forall\ s \in [K]\]
The algorithm's performance is compared to the \texttt{DYNAMIC} benchmark, which is the minimum cost solution that is supported by at least one expert at each step, formally:
\[\texttt{DYNAMIC} = \min_{\hat{\textbf{x}} \in \hat{X}} \sum_{i=1}^{n} c_i \hat{x}_i \textnormal{, where}\]
\[\hat{X} = \{\hat{\textbf{x}} : \forall\ i \in [n],\ \forall\ j \in [m],\ \exists\ s \in [K] \textnormal{ where the solution } x_i(j,s) \le \hat{x}_i \}\]
While a constraint is not satisfied, their algorithm updates each variable with an increasing rate of
\[\frac{dx_i}{dt} = \frac{a_{ij}}{c_i}(x_i + \delta_{ij})\]
where $\delta_{ij} = \frac{1}{K} \sum_{s=1}^{K} x_i(j,s)$ is the average of the experts' solutions for $x_i$ at the arrival of constraint $j$.
Algorithm 1 of \cite{AnandGe22:Online-Algorithms} scales down the problem with $0.5$, so it does not increase any variable above $0.5$ and satisfies each constraint with value $0.5$. The exact solution is obtained by doubling the variables at the end of the execution. (This descaling is an important aspect in the authors' proof.)

\subsection{Counter example}

In the following example we reveal in an online manner a linear program parametrized by $L$ with $K$ experts and observe the behavior of Algorithm $1$ (from \cite{AnandGe22:Online-Algorithms}).

\medskip

\noindent \textbf{Objective}. The example has $(L \cdot K + 1)$ variables with uniform cost:
\[ \min\ x_1 + x_2 + \dots + x_{K} + \dots + x_{2K} + \dots + x_{LK} + x_{LK+1}\]

\noindent \textbf{Constraints}. There are $L$ batches of $(K - 1)$ constraints. The first constraint of each batch has $(K+1)$ variables. The last variable ($x_{LK+1}$) is present in every constraint in every batch, but none of the experts suggests to use this variable. Within a batch, each consecutive constraint has one less variable. The experts set each variable that appears in later batches to $0$. The first batch:
\begin{align*}
     & \ \ \ \ x_{1} + x_{2} + \dots + x_{(K-1)} + x_{K} + x_{LK+1} \ge 1\\
\textnormal{Expert}_{1}: & \ \ \ \ \ 1 \ \ \ \ \ 0 \ \ \ \ \dots \ \ \ \ 0  \ \ \ \ \ \ \ \ \ \ \ 0 \ \ \ \ \ \ 0 \\
\textnormal{Expert}_{2}: & \ \ \ \ \ 0 \ \ \ \ \ 1 \ \ \ \ \dots \ \ \ \ 0  \ \ \ \ \ \ \ \ \ \ \ 0 \ \ \ \ \ \ 0 \\
     \vdots  & \\
\textnormal{Expert}_{K-1}: & \ \ \ \ \ 0 \ \ \ \ \ 0 \ \ \ \ \dots \ \ \ \ 1  \ \ \ \ \ \ \ \ \ \ \ 0 \ \ \ \ \ \ 0 \\
\textnormal{Expert}_{K}: & \ \ \ \ \ 0 \ \ \ \ \ 0 \ \ \ \ \dots \ \ \ \ 0  \ \ \ \ \ \ \ \ \ \ \ 1 \ \ \ \ \ \ 0 \\
     & \ \ \ \ \ \ \ \ \ \ x_{2} + \dots + x_{(K-1)} + x_{K} + x_{LK+1} \ge 1\\
\textnormal{Expert}_{1}: & \ \ \ \ \ 1 \ \ \ \ \ 1 \ \ \ \ \dots \ \ \ \ 0  \ \ \ \ \ \ \ \ \ \ \ 0 \ \ \ \ \ \ 0 \\
\textnormal{Expert}_{2}: & \ \ \ \ \ 0 \ \ \ \ \ 1 \ \ \ \ \dots \ \ \ \ 0  \ \ \ \ \ \ \ \ \ \ \ 0 \ \ \ \ \ \ 0 \\
     \vdots  & \\
\textnormal{Expert}_{K-1}: & \ \ \ \ \ 0 \ \ \ \ \ 0 \ \ \ \ \dots \ \ \ \ 1  \ \ \ \ \ \ \ \ \ \ \ 0 \ \ \ \ \ \ 0 \\
\textnormal{Expert}_{K}: & \ \ \ \ \ 0 \ \ \ \ \ 0 \ \ \ \ \dots \ \ \ \ 0  \ \ \ \ \ \ \ \ \ \ \ 1 \ \ \ \ \ \ 0 \\
     \vdots  & \\
     & \ \ \ \ \ \ \ \ \ \ \ \ \ \ \ \ \ \ \ \ \ \ \ \ x_{(K-1)} + x_{K} + x_{LK+1} \ge 1\\
\textnormal{Expert}_{1}: & \ \ \ \ \ 1 \ \ \ \ \ 1 \ \ \ \ \dots \ \ \ \ 1  \ \ \ \ \ \ \ \ \ \ \ 0 \ \ \ \ \ \ 0 \\
\textnormal{Expert}_{2}: & \ \ \ \ \ 0 \ \ \ \ \ 1 \ \ \ \ \dots \ \ \ \ 1  \ \ \ \ \ \ \ \ \ \ \ 0 \ \ \ \ \ \ 0 \\
     \vdots  & \\
\textnormal{Expert}_{K-1}: & \ \ \ \ \ 0 \ \ \ \ \ 0 \ \ \ \ \dots \ \ \ \ 1  \ \ \ \ \ \ \ \ \ \ \ 0 \ \ \ \ \ \ 0 \\
\textnormal{Expert}_{K}: & \ \ \ \ \ 0 \ \ \ \ \ 0 \ \ \ \ \dots \ \ \ \ 0  \ \ \ \ \ \ \ \ \ \ \ 1 \ \ \ \ \ \ 0 \\
\end{align*}

During the first constraint of every batch, the experts' solutions form an identity matrix. With each disappearing variable in the consecutive constraints, experts who suggested to use variables which are no longer available, choose to set the variable with the smallest index. Consequently, $(K-1)$ experts suggest to use variable $x_{(K-1)}$ and one expert suggests to use $x_K$ during the last constraint in the first batch. The pattern of the experts' solutions are identical for each batch. The constraints of the $l^{th}$ batch ($ 1 \le l \le L)$ are:
\begin{align*}
     x_{(l-1) K + 1} + x_{(l-1) K + 2} + \dots + x_{(l-1) K + (K-1)} + x_{lK} + x_{LK+1} \ge & \ 1\\
     x_{(l-1) K + 2} + \dots + x_{(l-1) K + (K-1)} + x_{lK} + x_{LK+1} \ge & \ 1\\
     \vdots &\\
     x_{(l-1) K + (K-1)} + x_{lK} + x_{LK+1} \ge & \ 1\\
\end{align*}

\begin{claim}
The objective value of Algorithm $1$ (from \cite{AnandGe22:Online-Algorithms}) on our example is $O(L \log(K))$ times the \texttt{DYNAMIC} benchmark.
\end{claim}
\begin{proof}
The optimal solution $\vect{x}^{*}$ of the \texttt{DYNAMIC} benchmark is the solution in which $x^{*}_{LK+1} = 1$ and $x^{*}_{i} = 0$ for $i \neq LK + 1$.
We verify that $\vect{x}^{*} \in \hat{X}$. For each $i \neq l K$ where $1 \leq l \leq L$, for each constraint $m$, $x^{*}_{i} \geq 0 = x_{i}(m,K)$.
For $i = l K$  where $1 \leq l \leq L$, for each constraint $m$, $x^{*}_{i} \geq 0 = x_{i}(m,1) = x_{i}(m,2) = \ldots = x_{i}(m,K-1)$.
Moreover, $\vect{x}^{*}$ satisfies all constraints (since variable $x_{LK+1}$ appears in all constraints).
Hence, $\vect{x}^{*} \in \hat{X}$. Subsequently, the objective value of the \texttt{DYNAMIC} benchmark is $1$.

     By the design of Algorithm $1$, the increasing rate of $x_{LK+1}$ is zero throughout the execution, and the variables which are not part of the current constraint are not increased. During the first constraint of each batch, the increasing rate of the first $K$ variables in the batch is $1/K$, since the increasing rate of variable $x_i$ is $(x_i + \frac{1}{K} \sum_{s=1}^{K} x_i(j,s))$ and initially every variable is set to zero. At the second constraint, the increasing rate of the second variable in the batch is higher than the other variables' increasing rate, because the first expert also uses this variable in its solution. Therefore, the increasing rate of the second variable is $(x_{(l-1) K + 2} + 2/K)$, while the other remaining expert variables in the constraint have an increasing rate of $(x_i + 1/K)$. Following the same reasoning (apart from the first constraint in the batch), the variable with the smallest index in the constraint has a higher increasing rate, than the other variables. During the last constraint of each batch, the increasing rate of the last two remaining expert variables are
     $(x_{(l-1) K + (K-1)} + (K-1)/K)$ and $(x_{lK} + 1/K)$. Keeping the increasing rates and the constraint satisfaction in mind, we can lower bound the value of each variable:
     \begin{align*}
          \frac{1}{K} \ \le& \ x_{(l-1) K + 1} \\
          \frac{1}{K-1} \ \le& \ x_{(l-1) K + 2} \\
          \frac{1}{K-2} \ \le& \ x_{(l-1) K + 3} \\
          & \ \vdots \\
          \frac{1}{3} \ \le& \ x_{(l-1) K + (K-2)} \\
          \frac{1}{2} \ \le& \ x_{(l-1) K + (K-1)} \\
          \frac{1}{K} \ \le& \ x_{lK} \\
     \end{align*}

     Summing the terms together, we get that the objective value increases at least with $O(\log K)$ during each batch. There are $L$ batches, so the total cost of Algorithm $1$ is at least $O(L \log(K))$,
     while the total cost of the \texttt{DYNAMIC} benchmark is $1$, which concludes the proof.
\end{proof}

\subsection{Comparison}

In this specific counter-example, the \texttt{LIN-COMB} benchmark is equivalent to the static best-expert benchmark, i.e., the solution of Expert$_K$. The objective value of \texttt{LIN-COMB} is $L$ (since the optimal solution sets $x_{lK}$ variables for $1 \leq l \leq L$  to one and other variables to 0). In this counter-example, the objective value of our algorithm is $O(L\log K)$. Consequently, our proposed algorithm is $O(\log K)$ competitive in the \texttt{LIN-COMB} benchmark.

\end{document}